\documentclass[12pt,twoside]{article}   
\usepackage[super,sort,comma]{natbib}

\usepackage{fancyhdr}		

\newcommand{\captionv}[3]{\begin{center}\parbox{#1cm}{\caption[#2]{{\sf #3}}}
\end{center}}

\usepackage[section]{placeins}   %

\usepackage{graphicx}

\makeatletter \renewcommand\@biblabel[1]{$^{#1}$} \makeatother
\newcommand{\cen}[1]{\begin{center} #1 \end{center}}

\usepackage{hyperref}
\hypersetup{ colorlinks,
    citecolor=blue,
    filecolor=blue,
    linkcolor=blue,
    urlcolor=blue
}

\usepackage{xcolor}

\definecolor{gray}{rgb}{0.6,0.6,0.6}
\definecolor{red}{rgb}{0.85,0,0}
\definecolor{green}{rgb}{0,0.85,0}
\definecolor{blue}{rgb}{0,0,0.85}
\definecolor{beige}{rgb}{0.92,0.87,0.78}
\usepackage[all]{hypcap}    

\begin{document}

\cen{\sf {\Large {\bfseries FastCAT: Fast Cone Beam CT (CBCT) Simulation} \\  
\vspace*{10mm}
Jericho O'Connell and Magdalena Bazalova-Carter} \\
University of Victoria,  3800 Finnerty Road, Canada
\vspace{5mm}\\
Version typeset \today\\
}

\pagenumbering{roman}
\setcounter{page}{1}
\pagestyle{plain}
Author to whom correspondence should be addressed. email: jerichoo@uvic.ca\\

\begin{abstract}
\noindent {\bf Purpose:} To develop fastCAT, a fast cone-beam computed tomography (CBCT) simulator. The fastCAT application uses pre-calculated Monte Carlo (MC) CBCT phantom-specific scatter and detector response functions to reduce simulation time for megavoltage (MV) and kilovoltage (kV) CBCT imaging. \\
{\bf Methods:} Pre-calculated x-ray beam energy spectra, detector optical spread functions and energy deposition, and phantom scatter kernels are combined with GPU raytracing to produce CBCT volumes. MV x-ray beam spectra are simulated with EGSnrc for 2.5 and 6 MeV electron beams incident on a variety of target materials and kV x-ray beam spectra are calculated analytically for an x-ray tube with a tungsten anode. Detectors were modelled in Geant4 extended by Topas and included optical transport in the scintillators. Two MV detectors were modelled, a standard Varian AS1200 GOS detector and a novel CWO high detective quantum efficiency detector. A kV CsI detector was also modelled. Energy dependent scatter kernels were created in Topas for two 16-cm diameter phantoms: A Catphan 515 contrast phantom and an anthropomorphic head phantom. The Catphan phantom contained inserts of 1-5 mm in diameter of six different tissue types; brain, deflated lung, compact and cortical bone, adipose and B-100.  \\
{\bf Results:} FastCAT simulations retain high fidelity to measurements and MC simulations: MTF curves were within 3.5\% and 1.2\% of measured values for the CWO and GOS detectors, respectively. HU values and CNR in a fastCAT Catphan 515 simulation were seen to be within 95 \% confidence intervals of an equivalent MC simulation for all of the tissues with root mean squared errors less than 16 HU and 1.6 in HU values and CNR comparisons, respectively. The anthropomorphic head phantom kV CBCT image resulted in a higher tissue contrast and a lower noise compared to the MV CBCT image. A fastCAT simulation of the Catphan 515 module with an image size of 1024$\times$1024$\times$10 voxels took 61 seconds on a GPU while the equivalent Topas MC was estimated to take more than 0.3 CPU years. \\
{\bf Conclusions:} We present an open source fast CBCT simulation with high fidelity to MC simulations. The fastCAT application can be found at \href{https://github.com/jerichooconnell/fastCATs.git }{https://github.com/jerichooconnell/fastCAT.git}.\\


\end{abstract}

\newpage     

\tableofcontents

\newpage

\setlength{\baselineskip}{0.7cm}      

\pagenumbering{arabic}
\setcounter{page}{1}
\pagestyle{fancy}
\section{Introduction}

Cone beam computed tomography (CBCT) is applied extensively in clinical radiotherapy. CBCT enables image-guided radiotherapy (IGRT) for the verification of patient positioning \cite{Grzadziel2007EPIDVerification}, quality assurance of treatment plans \cite{Mijnheer2013InRadiotherapy,Celi2016EPIDResults}, and real time tumor localization \cite{Grzadziel2007EPIDVerification}. Two different methods of CBCT are clinically applied, both of which include a source and detector mounted on a rotating gantry. The most common CBCT system uses a gantry mounted kilovoltage (kV) x-ray source and a detector. This is known as on-board imaging or kV-CBCT imaging. Detectors in kV-CBCT can be thin and made of low Z materials since the beams are not very penetrating. Conversely, a megavoltage (MV) treatment beam can be used as the source in MV-CBCT imaging. MV detectors, also known as electronic portal imaging devices (EPIDs), are generally made of higher Z materials and are thicker to stop high energy x-rays. kV-CBCT has better soft tissue contrast and higher image quality than MV-CBCT \cite{Alaei2015ImagingTherapy}. However, kV-CBCT requires an additional source and detector. Conversely, MV-CBCT generally has reduced image quality and is used in specific applications such as high-Z material artifact suppression \cite{Lindsay2019InvestigationDetector,Wu2014MetalImaging} but does not require an additional source. Performance of these imaging modalities is dependent on the x-ray source spectrum and detector design. As such, EPID development and imaging beam optimization are active areas of research. To optimize new imaging techniques, Monte Carlo (MC) simulation methods are often used. These techniques enable an initial evaluation of imaging setups before creating an expensive prototype of the detector or source.

MC EPID simulations are accurate but exceptionally computationally demanding. A large number of particles need to be simulated to form a CBCT image. The long simulation time can be primarily attributed to two factors. First, MV detectors generally have low detective quantum efficiency (DQE), meaning that many particles that are transported in the simulation do not interact with the detector and do not contribute to image formation. A typical EPID DQE is as low as 1\%-1.5\% \cite{Myronakis2017AOptimization:,Hu2017APerformance}. Second, the scintillator in which the optical photons are produced has a high scintillation yield; generating thousands of optical photons to be transported per interaction event. These two factors result in simulation times often as long as 3,000 core-hours for a 10$^7$ primary x-ray simulation of one EPID projection \cite{Blake2013CharacterizationGeant4}. Further, to produce a clinically equivalent image of 1 MU with a 10$\times$10cm$^2$ field size a simulation with more than 10$^{11}$ photons is required\cite{Shi2019ADetectors.,Star-Lack2014RapidDQEf,Rottmann2016AEfficiency}. This problem is compounded in CBCT simulation, where it is necessary to simulate many projections of the object for CBCT image reconstruction.

Large headway has been made to reduce this computational overhead. Star-lack \textit{et al.}  have shown that one can simulate the detector response with only a fraction of the scintillation yield, significantly reducing the computation time\cite{Star-Lack2014RapidDQEf}. Likewise, by simulating the optical spread function at discreet energies beforehand and convolving these optical spread functions with the energy deposition of an absorbed photon one can avoid simulating the scintillation processes completely\cite{Kausch1999MonteRadiotherapy, Kirkby2005ComprehensiveEPID}. Additionally, Shi \textit{et al.} introduced the fastEPID framework which pre-calculates energy deposition efficiency ($\eta$) and optical spread function (OSF) to remove particle transport in the detector entirely without loss of image quality \cite{Shi2019ADetectors.}. Where $\eta$ is defined as the ratio of the total energy deposition in the scintillator and the total x-ray photon energy incident on the detector. However, these simulations are still considered computationally intensive with one image at 1 MU taking 1.540 $\times$ 10$^4$ core-hours on an Intel Skylake CPU core (Intel Corp., Santa Clara, Ca).

Additionally, a number of works simulate fan and cone beam CT analytically to reduce computation time. ImaSim analytically simulates fan and cone beam CT through raytracing using vectorized phantoms \cite{Landry2013ImaSimRadiology}. VOXSI simulates kV fan beam CT using analytical raytracing with voxelized phantoms \cite{vanderHeyden2018VOXSI:Imaging} and shows agreement with experimental images in terms of image contrast. DukeSim simulates kV fan beam CT with voxelized phantoms through a combination of analytical raytracing and GPU MC and demonstrates agreement between experimental and simulated images in terms of image contrast, noise magnitude, noise texture, and spatial resolution \cite{Abadi2019DukeSim:Tomography}. A drawback to the DukeSim approach is that the GPU MC reduces the simulation speed, requiring a 2-3 minute simulation per source rotation while running on 4 Nvidia Titan Xp GPUs with 64 GB of memory. Overall, none of these platforms are open-source and only DukeSim shows agreement with experimental noise and image contrast. Additionally, none of these platforms show experimental agreement for kV or MV CBCT.

In this work we extend the fastEPID pre-calculation of MC data by pre-calculating the energy spectrum of the beam source as well as energy dependent scatter kernels for a cylindrical water phantom. We combine this data with an analytical GPU raytracer that provides the primary particle attenuation. This simulation strategy is used to create fastCAT, an open source simulation application for CBCT image formation to enable studies of novel beam, detector, and phantom combinations. FastCAT shows good agreement with MC simulations of full CBCT data acquisition and it results in extremely short run times on the order of 1 GPU-minutes for a full CBCT simulation.

\section{Materials and Methods}

\begin{figure}[ht!]
   \begin{center}
   \includegraphics[width=\textwidth,trim={3cm 4cm 2.8cm 4cm}, clip]{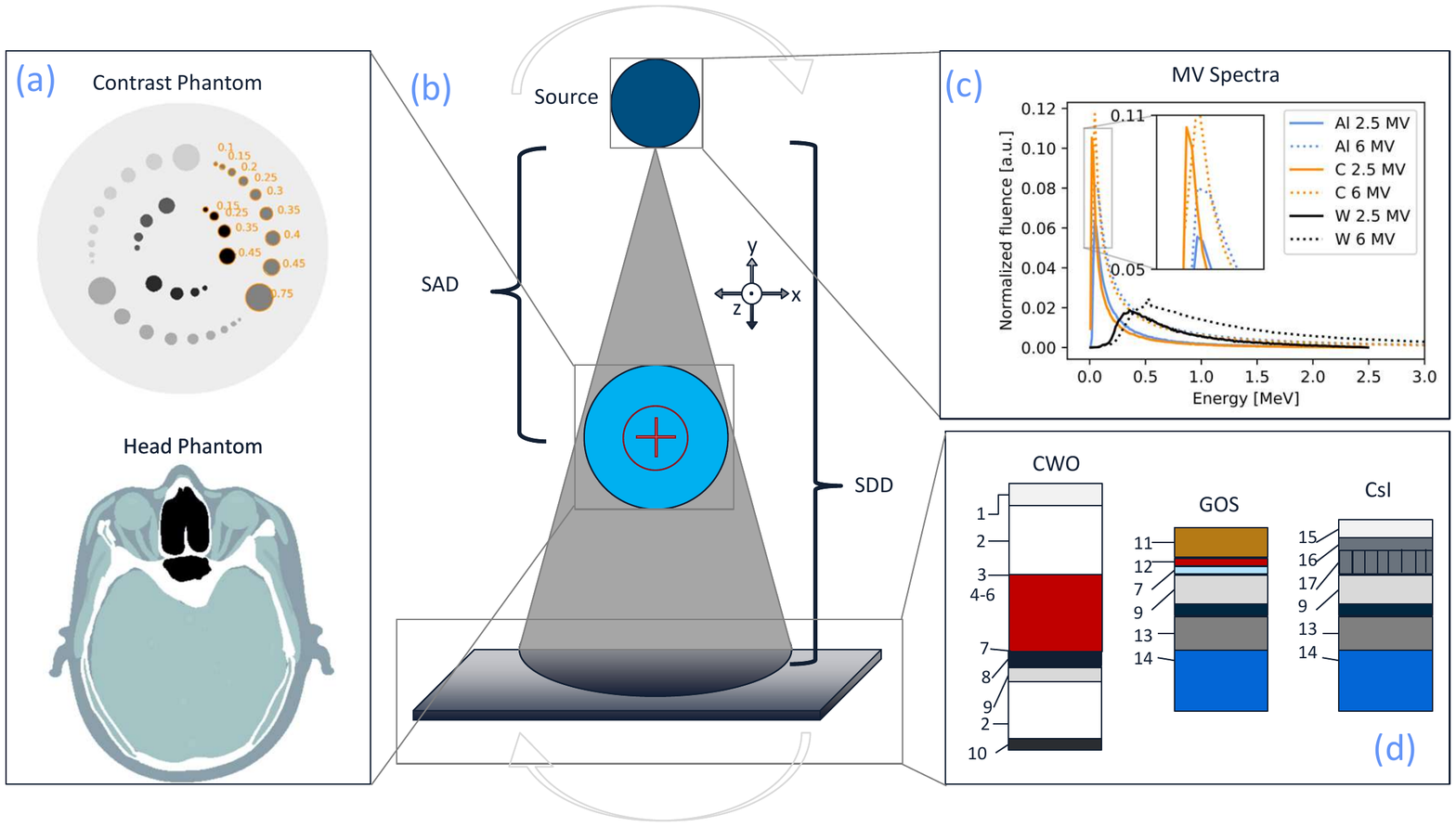}
   \captionv{12}{}
   {An overview of the simulation setup. a) The two phantoms used in this work, the Catphan 515 (top) and the XCAT anthropomorphic head phantom (bottom). b) The simulation geometry. c) The MV energy spectra. d) A slice of the CWO (left), GOS (center), and CsI (right) detectors with component labels corresponding to Table \ref{tab:detector parameters}.
   \label{fig_setup} 
    }  
    \end{center}
\end{figure}

\begin{figure}[ht]
   \begin{center}
   \includegraphics[width=\textwidth,trim={0 6cm 4cm 3cm},clip]{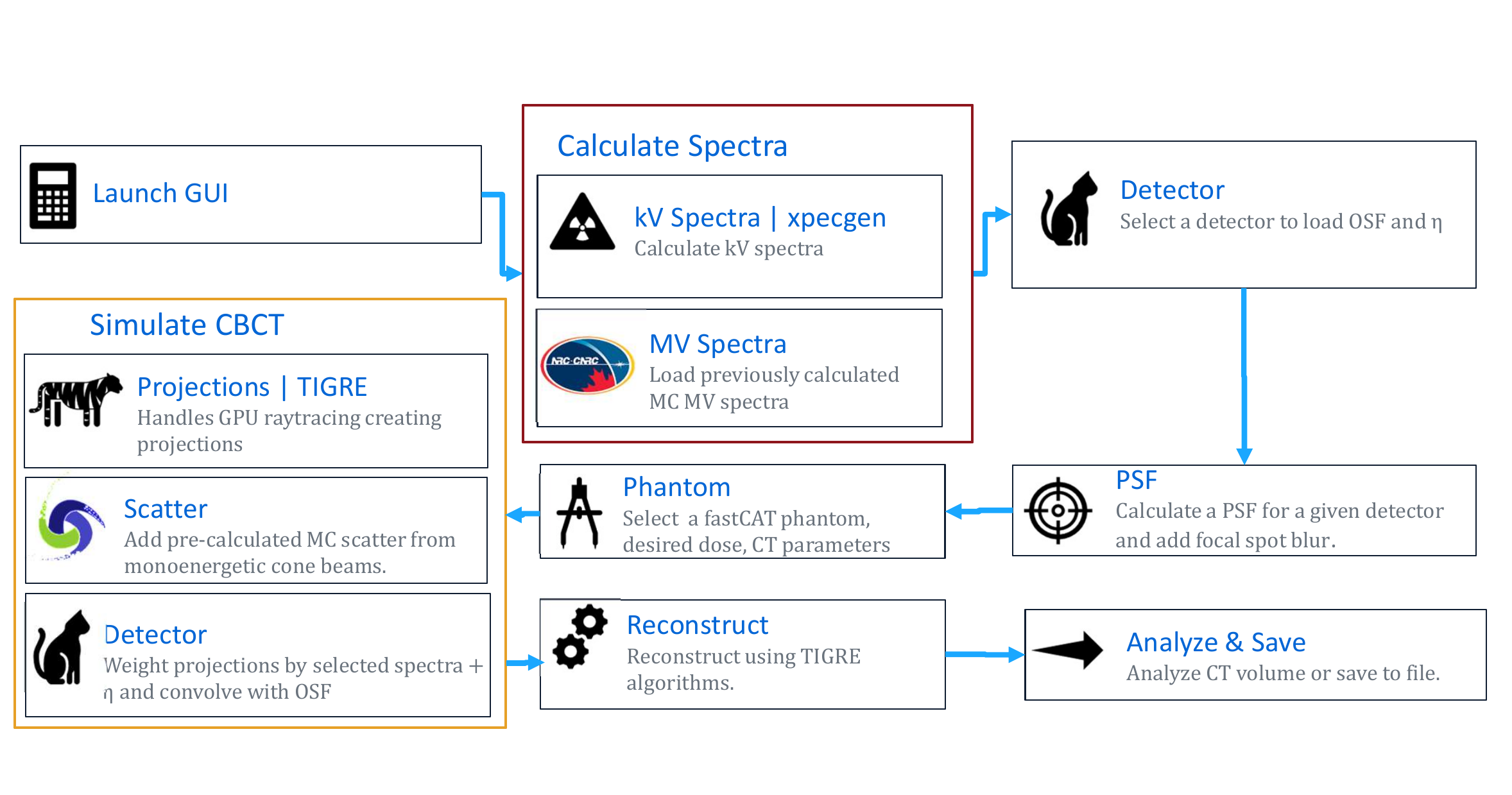}
   \captionv{12}{Short title - can be blank}
   {An overview of the fastCAT workflow.
   \label{fig_overview} 
    }  
    \end{center}
\end{figure}

The fastCAT applications consists of a number of components that are described below. The fastCAT graphical user interface (GUI) is built on top of xpecgen \cite{Hernandez2016Xpecgen:Anodes}, an open source x-ray tube spectrum generator written in python (version 3.6). The fastCAT default simulation geometry is depicted in Figure \ref{fig_setup}: A cone beam collimated to 16$\times$16 cm$^2$ at isocenter impinges on a 16 cm diameter cylindrical phantom at a source-to-axis distance (SAD) of 1.00 m and a source-to-detector distance (SSD) of 1.52 m. The GUI workflow is depicted in Figure \ref{fig_overview}, which also the user selected parameters.

\subsection{Imaging Beam}

FastCAT can use MV or kV x-ray beams for CBCT image generation. MV spectra were calculated in EGSnrc/BEAMnrc \cite{IKawrakow2018TheTransport} using a variety of target materials (carbon, aluminum, and tungsten) and electron beam energies (2.5 and 6 MeV) (Figure \ref{fig_setup}c). To generate the spectra a mono-energetic 0.02 mm diameter electron beam was incident on the target material. The thicknesses of the target materials were based on existing experimental beams for the carbon and aluminum targets \cite{Flampouri2002OptimizationExperiment} and are summarized in Table \ref{tab:targets}. For the low atomic number ($Z$) beams (carbon and aluminum), a 2 cm polystyrene filter placed at 50 cm from the target was used to reduce electron fluence in the beam. The photons were collimated by both primary and secondary collimators to create a 10$\times$10 cm$^2$ field size at isocenter and phasespace files were scored at a distance of 100 cm from the target. Global electron and photon cutoffs were set to 0.01 and 0.001 MeV in the BEAMnrc simulations. 1.5$\times$10$^8$ histories were used in each run resulting in 1$\times$10$^6$ to 3$\times$10$^7$ photons in the phasespace file. These simulations took on average 35 core-hours on an Intel Skylake CPU.  Energy spectra with 200 evenly spaced energy bins between 0 and the maximum energy were generated from the phasespace files for input into fastCAT (Figure \ref{fig_setup}c). The kV spectrum was modelled for a 120 kVp beam with a 12$^\circ$ tungsten anode, a 0.4 mm beryllium window, and 1 mm aluminum filter in xpecgen \cite{Hernandez2016Xpecgen:Anodes}.

\begin{table}[ht]
\begin{center}
\caption{Electron target thicknesses used for the generation of the MV imaging beams.}
\vspace*{1ex}
\label{tab:targets}
\begin{tabular}{llll}
 \hline
       & tungsten & aluminum & carbon \\  \hline
2.5 MV & 2.3 mm   & 6.7 mm \cite{Parsons2014AVirtuaLinac} & 7.6 mm \cite{Parsons2012BeamTargets}\\
6 MV   & 5 mm     & 8 mm \cite{Baek2019AssessmentFilter}     & 9.9 mm \cite{Parsons2012BeamTargets}
\end{tabular}
\end{center}
\end{table}

\subsection{Detector Simulation \label{sec:det}}

FastCAT currently employs two MV detectors: a novel Cadmium Tungstate (CWO) detector\cite{Star-Lack2015AImaging}  and a Gadolinium Oxysulfide (GOS) AS1200 detector\cite{Shi2018APerformance} (both Varian Medical Systems, Palo Alto, CA).  A detector schematic is depicted in Figure \ref{fig_setup}d and detector material parameters are summarized in Table \ref{tab:detector parameters}.  The CWO detector has pixel septa between crystals and has a pixel pitch of 0.784 mm. The GOS detector has no septa and pixel pitches are set by the active matrix flat-panel imager (AMFPI) pixel pitch. AMFPI pixel pitches of 0.784, 0.392 or 0.336 mm were modelled for the GOS detector and 0.784 and 0.392 mm were modelled for the CWO detector. 
The response of the three detectors, including the optical transport, was simulated in Geant4 \cite{Agostinelli2003Geant4Toolkit} using the Topas \cite{Perl2012Topas:Applications} wrapper.
The responses of the detectors were modelled using a modification of the fastEPID framework developed by Shi \textit{et al.}\cite{Shi2019ADetectors.}. Mono-energetic pencil beams of energies of 10 to 90 keV in 10 keV increments, 100 to 900 keV in 100 keV increments, and 1, 2, 4, and 6 MeV were used to calculate the optical spread function (OSF) as well as the energy deposition efficiency ($\eta$) of each detector. For each of the energies, 1$\times$10$^6$ histories were simulated resulting in a run time of roughly 48 core-hours per detector on an Intel Skylake CPU for the GOS detector. The other detectors had comparable run times but took longer to load the complex CWO pixels and columnar CsI geometries, which took 2 and 6 hours to load, respectively. All Topas simulations used the Geant4 Penelope physics list as well as the Geant4 optical physics list with a particle range cutoff of 5 $\mu m$ for all particles. No variance reduction techniques were used.

 The kV CsI detector in this study was modelled as a 450-$\mu$m CsI scintillating detector\cite{Sharma2012EffectiveGlasses} (Radiation Monitoring Devices, Inc., Water-town MA).  The scintillating crystal is composed of a 100 $\mu$m graphite substrate, 90 $\mu$m of homogeneous CsI, and 360 $\mu$m of columnar CsI (Figure \ref{fig_setup}d). The columnar CsI was simulated with a column diameter of 10.2 $\mu$m and an 85\% packing density with nitrogen gas between columns. Optical properties are summarized in Table \ref{tab:detector parameters}. All MC simulation settings were the same as stated above for the CWO and GOS detectors. To reduce computation time in all detector simulations, scintillation yields of 600 photons per MeV were used.

\begin{table}[]
\caption{Detector optical parameters
\cite{Star-Lack2015AImaging,Shi2018APerformance,Freed2009ExperimentalScreens}}
\label{tab:detector parameters}
\label{tab:detector parameters}
\begin{tabular}{llllll}
\hline
                       & \begin{tabular}[c]{@{}l@{}}Density\\ (g/$cm^3$)\end{tabular} & Material                & \begin{tabular}[c]{@{}l@{}}Thickness\\ (mm)\end{tabular} & \begin{tabular}[c]{@{}l@{}}Abs. length (cm) $|$\\ Refr. ind.\end{tabular} & Reflectivity \\ \hline
(1) carbon Fiber       & 1.62                                                         & C                       & 2.5                                                      & --                                                                              & --           \\
(2) Foam               & 0.05                                                         & C                       & 33.1, 25.0                                                & --                                                                              & --           \\
(3) Vikuiti ESR        & 1.05                                                         & CH                      & 0.065                                                    & 0.01                                                                            & 0.98         \\
(4) Scintillator Pixel & 7.9                                                          & CdWO$_4$                & 15                                                       & 125 $|$ 2.25                                                                    & --           \\
(5) Pixel Glue         & 1.0                                                          & Epoxy                   & 15                                                       & 100 $|$ 1.47                                                                    & --           \\
(6) Pixel Septa        & 2.7                                                          & Al Mylar                & 15                                                       & 0.001                                                                           & 0.88         \\
(7) Meltmount Glue     & 1.0                                                          & C$_{21}$H$_{25}$ClO$_5$ & 0.01                                                     & 300 $|$ 1.7                                                                     & --           \\
(8) Mylar              & 1.38                                                         & C$_{10}$H$_8$O$_4$      & 0.065                                                    & 100 $|$ 1.65                                                                    & --           \\
(9) AMFPI              & 2.6                                                          & SiO$_2$                 & 1                                                        & 0.001 $|$ 1.70                                                                  & --           \\
(10) Fiberglass        & 1.85                                                         & SiO$_2$                 & 0.6, 6.0                                                  & --                                                                              & --           \\
(11) Copper buildup    & 8.9                                                          & Cu                      & 1                                                        & --                                                                              & --           \\
(12) GOS phospor       & 4.59                                                         & Gd$_2$O$_2$S:Tb         & 0.29                                                     & 43 $|$ 2.3, 1.0 (binder)                                                        & --           \\
(13) Al alloy          & 2.8                                                          & Al                      & 1                                                        & --                                                                              & --           \\
(14) Pb alloy          & 10.95                                                        & Pb                      & 3                                                        & --                                                                              & --           \\
(15) Graphite          & 2.26                                                         & C                       & 1                                                        & 0.001                                                                           & 0.88         \\
(16) CsI               & 4.51                                                         & CsI:Tl                  & 0.9                                                      & 1.25 $|$ 1.8                                                                    & --           \\
(17) Columnar CsI      & 4.51                                                         & CsI:Tl                  & 3.60                                                     & 1.25 $|$ 1.8                                                                    & --          
\end{tabular}
\centering
\vspace*{2ex}
\end{table}

As stated above, Topas MC simulations were used to determine the OSF and $\eta$ of each detector. The OSF is defined as

\begin{equation}
    OSF(i,j) = \frac{Output(i,j)}{N_{incident}\times\eta},
\end{equation}

\noindent where $Output(i,j)$ is the number of optical photons incident on the readout pixel with row and column indices of $i$ and $j$ and $N_{incident}$ is the total number of photons incident on the detector.

In this work, the OSFs were used to generate a detector modulation transfer function (MTF) from the point spread function (PSF). The 2D energy dependent OSFs were weighted by energy deposition efficiency multiplied with the incident spectrum from the selected beam to estimate the hypothetical PSF. This PSF was then convolved with an idealized 0.3 mm wide slit tilted at 2.5$^\circ$ to generate a line spread function (LSF), that was then presampled to estimate the MTF. The OSF was also used in the fastCAT CBCT simulation to generate the detector response, as discussed in section II.C. 

\subsection{Imaging Phantoms}

 Two fastCAT phantoms are demonstrated in this work (Figure \ref{fig_setup}a): a Catphan 515 module with modified material compositions (The Phantom Laboratory, Salem NY) and the head of the anthropomorphic XCAT phantom \cite{Segars20104DResearch}. All phantoms in fastCAT are saved as numpy integer arrays containing integers between 0 and $n - 1$, where $n$ is the number of materials in the phantom. A corresponding phantom map with a string array of $n$ material names from the fastCAT material database is input into the fastCAT application. The fastCAT material database contains 336 materials from the NIST XCOM database\cite{BergerXCOM:1.5}, Geant4 default materials\cite{Agostinelli2003Geant4Toolkit}, and select materials from ICRU-44\cite{White1989Report44}. Additional materials can be uploaded as a csv file containing energy and linear attenuation coefficient values.
 
 The Catphan 515 phantom contained 5 to 15-mm in diameter inserts filled with five contrast materials. The default contrast materials in the Catphan 515 phantom were deflated lung, compact and cortical bone, adipose, brain and B-100, with material composition as defined in Geant4 default materials.
 
 Additionally, the use of an XCAT phantom for CBCT imaging is demonstrated. Selected slices of the XCAT head phantom were first converted into a 1024$\times$1024$\times$10 binary array of attenuation coefficients with a voxel size of 0.5$\times$0.5$\times$3.125 mm$^3$, which was further converted into the fastCAT phantom format: The parameter files of the XCAT phantom were used to identify materials in the phantom by their attenuation coefficient. These identified materials were assigned equivalent fastCAT materials with energy dependent linear attenuation coefficients calculated using compositions from ICRU-44. The XCAT phantom is under the authors' individual license and not available in the fastCAT GitHub repository.

 \subsection{Full CBCT Topas Simulations}
 
 Topas was used to validate fastCAT by means of a full CBCT imaging simulation as well as by means of projection images. A full CBCT image of the contrast phantom for a 6 MV tungsten beam was simulated. A cone beam collimated to 16$\times$16 cm$^2$ at isocenter impinged on the 16 cm diameter Catphan 515 phantom with SDD of 1.52 m and SAD of 1.00 m. Projection images of the Catphan 515 phantom were simulated at 180 evenly spaced angles distributed over a full 360$^\circ$ rotation. Particle counts were scored on an air slab at 1.52 m away from the source, and compared to fastCAT values. Simulation parameters described in the previous section were maintained. Particle transport through the full detector was not simulated as it resulted in a prohibitively long simulation time.  Note that the fastCAT detector simulation was validated separately through comparing detector MTF to experimental detectors (\ref{sec:det}). CBCT MC simulations of the Catphan 515 phantom were performed using $2\times10^8$ particles per projection image. Full CBCT simulations were run on a high-performance computing cluster using 180$\times$20 Intel E5-2683 v4 Broadwell cores. One full CBCT simulation took on average 37 core days. 
 
 Additional simulations of a single projection using other beam energy/target material combinations as well as average phantom dose were performed on a desktop computer using 8 Intel Skylake cores. The projection simulation parameters were the same as the CBCT simulations save the beam energy spectra, average phantom dose was scored simultaneously in these simulations. These simulations took on average 6 core-hours.
 
 \subsection{Scatter Correction}

The scatter correction for a 16-cm diameter water phantom was generated in Topas using 18 mono-energetic cone beams and the default simulation geometry. A phasespace file was collected at the surface of a 40$\times$10$\times$0.3 cm$^3$ air slab located at the 1.52 cm SDD. Photons that did not interact in the phantom were filtered out. The spatial distribution of the scattered particles $y$ was averaged in the $z$ direction and fitted to a two parameter curve-fit of the form:

\begin{equation}
    y = (\frac{a}{\sqrt{x^2 + a^2}})^b,
\end{equation}

where $a$ and $b$ are fitting parameters and $x$ is the off axis distance. This fit was used to ensure a symmetric fit, and it resulted in a root mean squared error lower than a symmetric polynomial fit. The resultant energy distribution of the scatter from each mono-energetic cone beam was neglected in the analysis as sufficiently close agreement was seen when approximating the scatter as mono-energetic. These mono-energetic cone beam scatter curves were then combined with analytical projections to form the CBCT projection data discussed below.

\subsection{Raytracing}

The bulk of the computational work of fastCAT CBCT image generation is in the CBCT raytracing. In this work the TIGRE \cite{Biguri2016TIGRE:Reconstruction} GPU reconstruction package was used for generating the cone beam forward projections. These primary projections, are denoted as “projections” and they are equivalent to the sum of attenuation coefficient along the path of the ray between the source and detector. This is to be distinguished from what we will call the “intensity profiles” which are the counts in a given detector pixel. 

One forward projection is calculated for each CBCT projection for each of the 18 energies. These primary projections are converted to intensity profiles using a flood field intensity profile made with the same geometry and number of photons as the scatter corrections, ensuring the correct scaling of the scatter. The scatter corrections are then added to the primary intensity profiles. These complete intensity profiles, $I(E)$, are then weighted by the fluence $\phi(E)$ and energy deposition efficiency $\eta(E)$ to form the final intensity $I_f$ as

\begin{equation}
    I_f = \sum\limits_{E} \phi(E) \eta(E) I(E)
\end{equation}

 To reduce computation time only ten slices in the z direction are simulated in the default phantoms, 512$\times$512$\times$10 voxels with voxel sizes of 0.31$\times$0.31$\times$0.31 mm$^3$ were used for the Catphan 515. Likewise, the XCAT phantom had dimensions of and 1024$\times$1024$\times$10 voxels with voxel sizes of 0.5$\times$0.5$\times$3.125 mm$^3$.

\subsection{Noise and CBCT reconstruction}

Noise is added through a Poisson scaling of the particle counts incident on the detector. This scaling can be related to either dose or particle fluence as will be discussed below. The intensity profiles are then convolved with the OSF at each energy and summed to create the final profile at each beam angle. These are converted back to projections using the corresponding flood field and reconstructed. Reconstructions are performed with TIGRE algorithms which have many options for reconstruction types including the default FDK reconstruction \cite{Feldkamp1984PracticalAlgorithm} and more advanced iterative reconstruction. CBCT images were converted into Hounsfield Units (HU) using water region of the phantom for normalization. Image analyses can then be done using specific analysis modules attached to each phantom. For the Catphan 515 module, contrasts and contrast to noise ratios can be calculated. CNR was calculated as 

\begin{equation}
    CNR = \frac{|HU_w - HU_{ROI}|}{\sqrt{\sigma_w^2 + \sigma_{ROI}^2}}
\end{equation}

\noindent Where $HU_w$ and $HU_{ROI}$ are the HU values water and the region of interest (ROI) and the regions respective standard deviations are represented by the $\sigma$s.

\subsection{Dose calculation}

The scaling of the noise in the simulation can be determined by the user based on either the requested mean dose to the phantom or the total particle fluence. Noise defined by particle fluence scales the intensity profile as $H$:

\begin{equation}
\label{nphot}
    H = \frac{N_\gamma}{A/A_p},
\end{equation}

where $N_\gamma$ is the total number of particles $A$ is the area of the beam and $A_p$ is the area of a detector pixel. Once $H$ is found the flood field is scaled to that height before converting the projections to intensity profiles. Scatter is then added using Poisson noise based on the number of counts in a pixel.  

Another option is to provide the mean dose to the water phantom for one projection. In this case the dose estimate per particle ($D_0$) at each energy is calculated from:

\begin{equation}
    D_0(E) = \frac{1}{N_i M}\sum\limits_i E \frac{\mu_{en}(E)}{\mu(E)} (1 - e^{-\mu_i(E) x_i}),
\end{equation}

\noindent where $i$ is the ray index, $\mu$ and $\mu_{en}$ are the linear and energy-absorption coefficients of water, respectively, $N_i$ is the number of rays, and $M$ is the mass of the phantom. These values estimate the average dose per particle at each energy. These doses are weighted by the selected fluence $\phi$ and summed to get a total dose estimate $D_p$ per particle

\begin{equation}
   D_p = \sum\limits_{E} \phi(E) D_0(E).
\end{equation}

\noindent These dose estimates were seen to correlate linearly (R$^2$ $>$ 0.99) with the mean dose to the 16 cm diameter water phantom calculated in Topas. An empirical linear fit to MC doses was used to relate these estimates to the final dose estimate per particle. 

The noise was then calculated based on the number of particles. This was achieved by dividing the requested dose by the dose per particle. With the number of particles known, equation (\ref{nphot}) was used to calculate the noise. The dose calculations were validated by comparing the average dose to the Catphan 515 phantom estimated in fastCAT to the dose calculated using the same imaging setup in Topas.

\section{Results}

\subsection{Detector MTF}

\begin{figure}[t!]
   \begin{center}
   \includegraphics[width=0.9\textwidth]{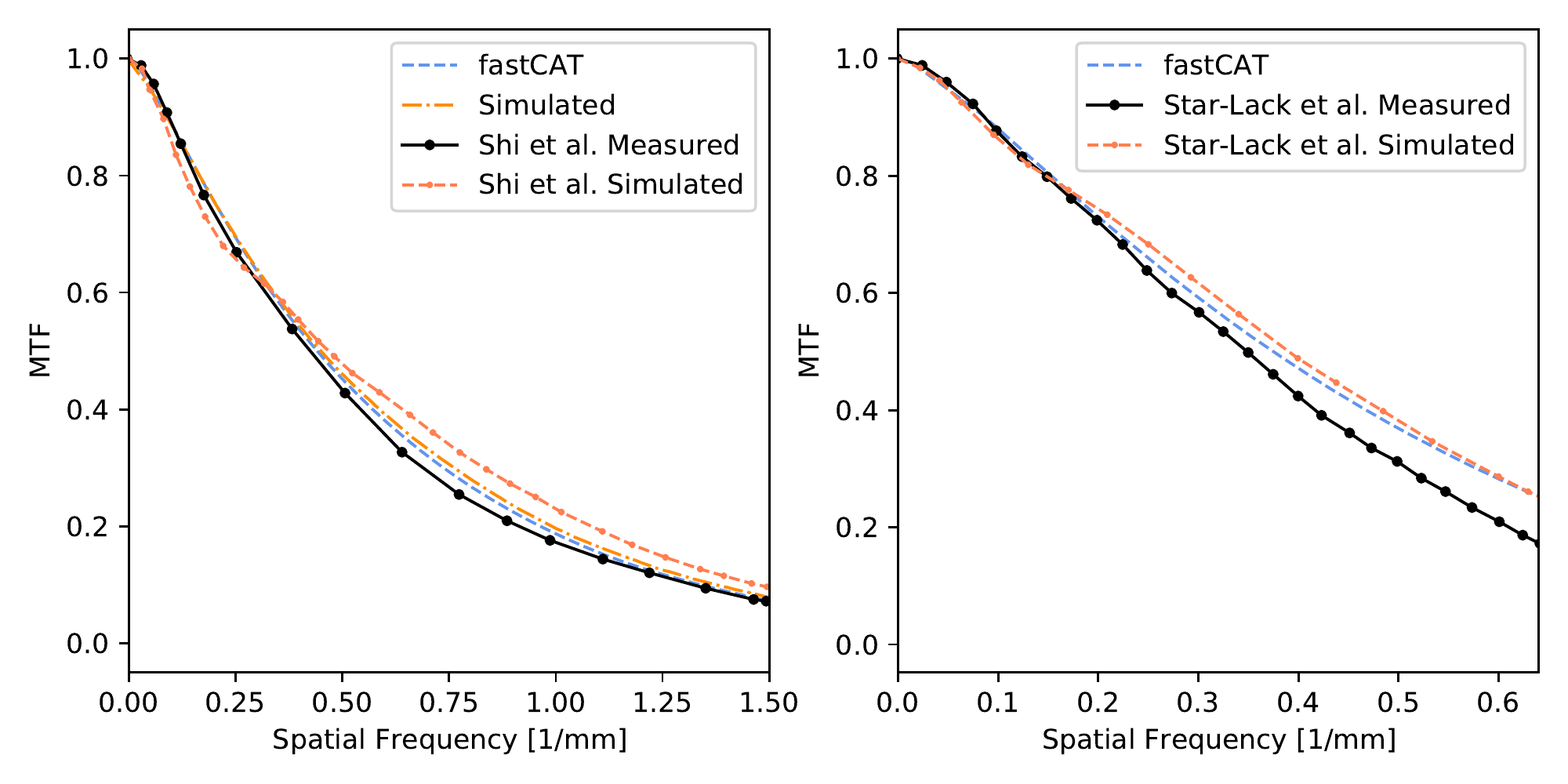}
   \captionv{12}{}
   {(L) A comparison of fastCAT-calculated MTF with measured and simulated values for the GOS detector with a 0.336 mm pixel pitch (based on Shi \textit{et al.})  and (R) for the CWO detector with a 0.784 mm pixel pitch (based on Star-lack \textit{et al.}).
   \label{MTF_comparison} 
    }  
    \end{center}
\end{figure}

FastCAT showed very good agreement with measured and simulated MTFs for both the GOS and CWO detectors (Figure \ref{MTF_comparison}). For the GOS detector the average root mean squared error (RMSE) between the fastCAT and Topas simulations was 0.5\%. The maximum RMSE was 1.3\% at an MTF of 0.61 mm$^{-1}$. The average RMSE and maximum RMSE between the GOS MTF calculated by fastCAT and measurement data provided by Shi \textit{et. al} was 1.2\% and 2.8\% at an MTF of 0.67 mm$^{-1}$, respectively. For the CWO detector, the average RMSE between the fastCAT and simulated data of Star-Lack \textit{et al.} was 3.3 \% and the largest discrepancy of  7.9 \% occured at a spatial frequency of 0.64 mm$^{-1}$. The average RMSE between the fastCAT and the measured data of Star-lack \textit{et al.} was 3.5 \% and the maximum RMSE of 7.9 \% was observed at a spatial frequency of 0.64 mm$^{-1}$.

\begin{figure}[ht!]
   \begin{center}
   \includegraphics[width=0.9\textwidth,trim={0 0 0 0},clip]{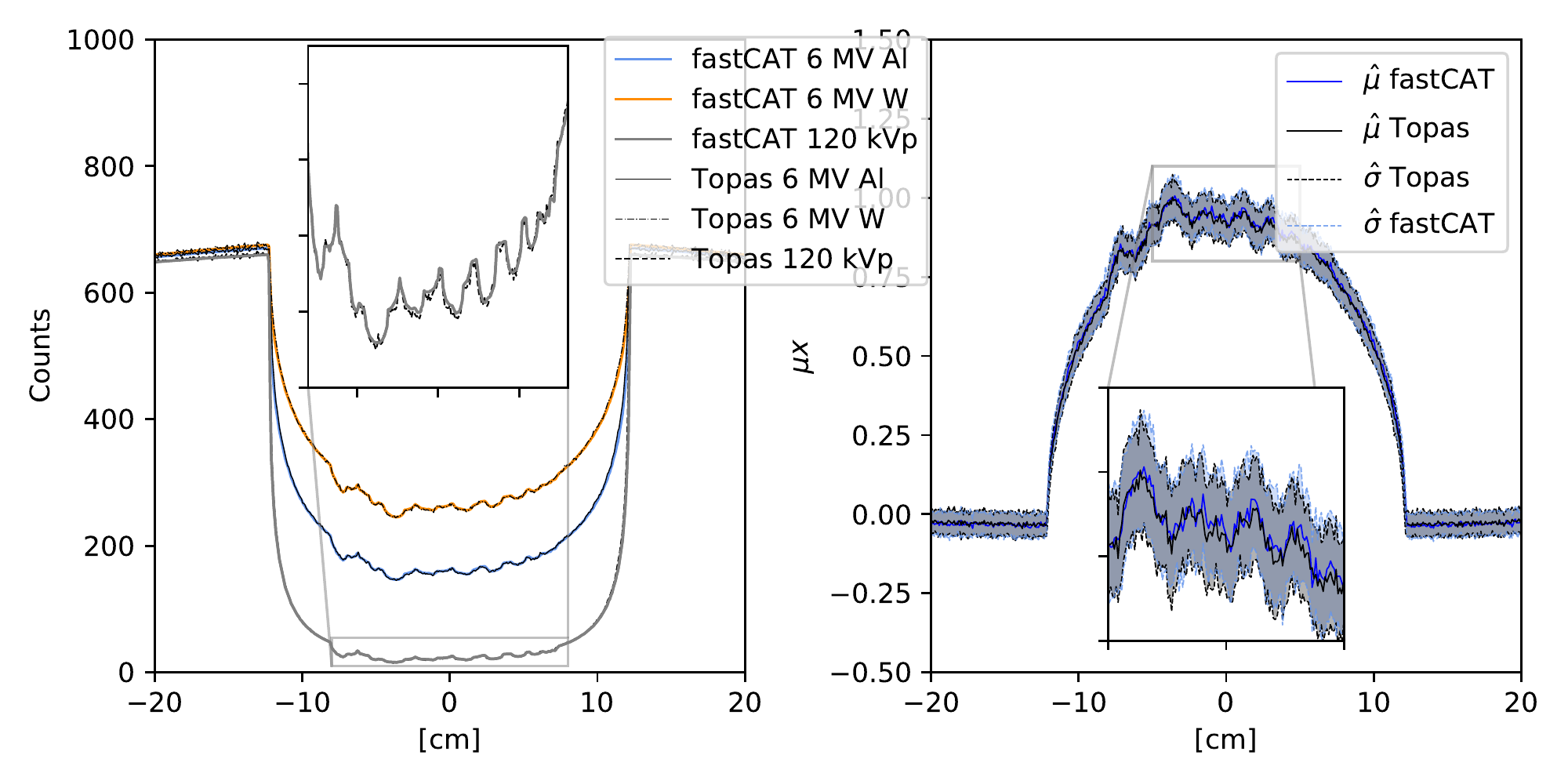}
   \captionv{12}{}
   {(L) Comparison of fastCAT and Topas intensity profiles for a 6 MV aluminum beam, a 6 MV tungsten beam, and a 120 kVp x-ray tube beam. (R) Comparison of fastCAT and Topas scatter profiles for a 6 MV tungsten projection (averaged over 64 pixels in the $z$ direction).
   \label{one_slice_comparison} 
    }  
    \end{center}
\end{figure}

\subsection{Projections and dose}

Comparisons of the Catphan 515 phantom projections and intensity profiles as calculated by fastCAT and Topas are shown in Figure \ref{one_slice_comparison}. The intensity profiles for the 6 MV aluminum beam presented in Figure \ref{one_slice_comparison} on the left demonstrated a close agreement between fastCAT and Topas. The fastCAT and Topas intensity profiles had an average RMSE of 0.4\% with a worst case RMSE of 1.1\%.  Likewise, for the 6 MV tungsten beam, fastCAT had an average RMSE of 0.2\% of the Topas values with a worst case RMSE of 1.4\%. The 120 kVp beam fastCAT  intensity profile had an average RMSE of 0.5\% of the Topas values with a worst case RMSE of 1.7\%. The lowest accuracy was likely due to noise in Topas simulations. FastCAT noise showed good agreement with MC noise as seen in Figure \ref{one_slice_comparison} on the right. The RMSE of the standard deviation of all pixels between fastCAT and Topas was 0.21 \% for the 6 MV tungsten beam.

FastCAT imaging dose calculations for the Catphan 515 phantom were compared to dose calculated for the same simulation in Topas. The mean dose to entire phantom was in a good agreement with the Topas dose estimates. FastCAT dose per photon had a mean difference of 1.4\% of the Topas values for all beams. The largest dose estimation error between fastCAT and Topas was for the 2.5 MV carbon beam with an error of 4.5 \%.

\begin{figure}[ht!]
   \begin{center}
   \includegraphics[width=0.9\textwidth]{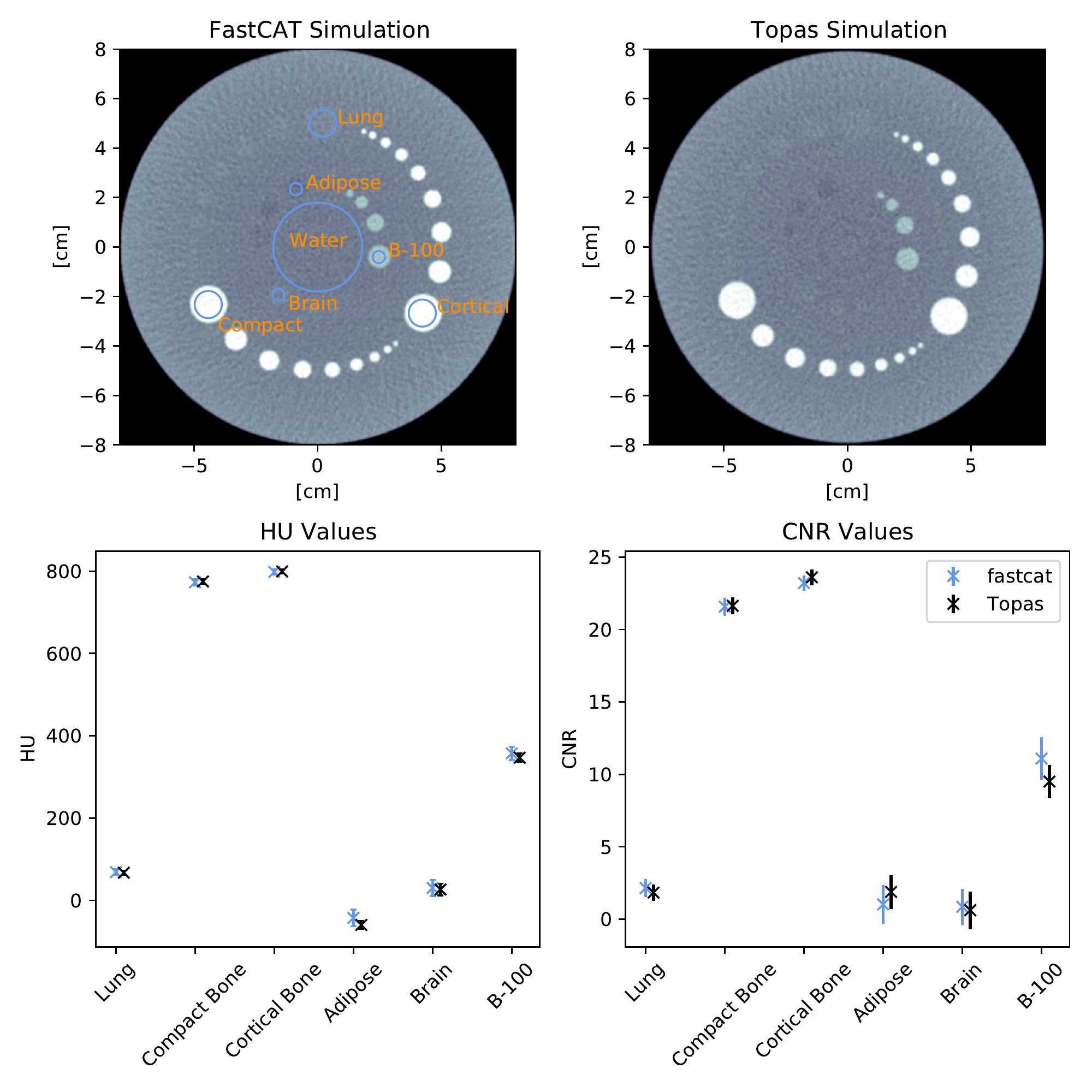}
   \captionv{12}{}
   {(Top) FastCAT simulation with ROI placement used for HU and CNR calculations (L) and full Topas simulation of the Catphan 515 phantom (R). (Bottom) A comparison of HU values (L) and CNR (R) in the CBCT fastCAT and Topas reconstruction.
   \label{recon_comparison} 
    }  
    \end{center}
\end{figure}

\subsection{CBCT images, CNR and calculation time}

The agreement in intensity profiles translated into good agreement between HU values in a full CBCT reconstruction of the Catphan 515 phantom (Figure \ref{recon_comparison}). The average RMSE between fastCAT and MC-calculated contrast was 0.5\%. The mean error for each of the inserts was 0.6, -1.4, -1.3, 15.8, 4.2, and 10.2 HU for the deflated lung, compact and cortical bone, adipose, brain and B-100, respectively. All errors were within the 95\% confidence interval.

The good contrast agreement also extended to good CNR agreement between fastCAT and Topas simulations (Figure \ref{recon_comparison}). The average RMSE between fastCAT and MC in terms of CNR was 0.55. The RMSE for the each of the inserts was 0.30, 0.08, 0.39, 0.79, 0.15 and 1.59 for the deflated lung, compact and cortical bone, adipose, brain and B-100, respectively. All errors were within the 95\% confidence interval.

\begin{figure}[h!]
   \begin{center}
   \includegraphics[width=0.8\textwidth,trim={0 0 0 0},clip]{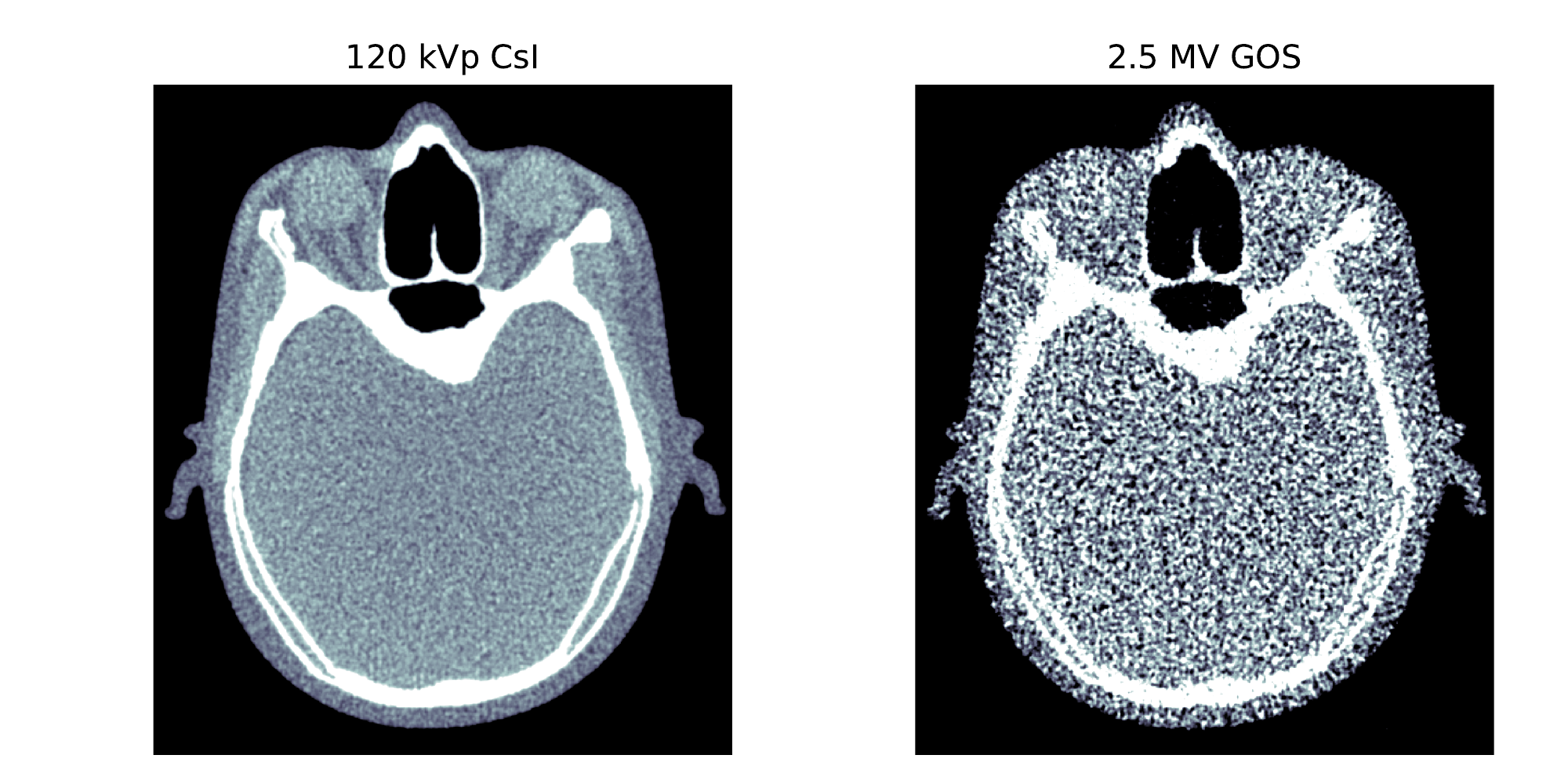}
  \captionv{12}{}
   {FastCAT CBCT images of an anthropomorphic head phantom created with (L) a 120 kVp beam and the CsI detector and (R) a 2.5 MV tungsten beam and the GOS detector. CBCTs were normalized to a mean dose to the phantom of 7 mGy and used 360 projections.
   \label{XCATs} 
    }  
    \end{center}
\end{figure}

To demonstrate fastCAT capability to simulate anthropomorphic phantoms, CBCT images of the head of the XCAT phantom are shown in (Figure \ref{XCATs}). As expected, the 120 kVp CBCT image simulated with the CsI detector is characterized by a lower noise and higher contrast compared to the 2.5 MV tungsten CBCT image acquired with the GOS detector.

A computational time comparison for CBCT simulations performed with a full Topas MC simulation, fastEPID and fastCAT for a 180-projection CBCT dataset generated with 6 MV tungsten beam was performed. An extrapolation is made for a full Topas MC simulation based on the time for single projection simulation with 10$^9$ photons. The extrapolation predicts 5.6 core-years of compute time to create a full CBCT. A fastEPID simulation using the same parameters would take an estimated 0.32 core-years, while the fastCAT simulations takes 40 and 61 seconds for a 512$\times$512$\times$10 and 1024$\times$1024$\times$10 reconstructed image size respectively.

\section{Discussion}
While the end goal of this platform is to achieve agreement of fastCAT and experimental CBCTs data, this work focuses on agreement between fastCAT simulations and MC simulations as an initial validation.
Overall, there was a close agreement between fastCAT and MC-generated MV CBCT images and the next step will consist of experimental validation. One validated fastCAT result is the MTF of the GOS and CWO detectors, which had a mean error of 1.2\% and 3.5\% of detector measurements performed by Shi \textit{et al.} and Star-lack \textit{et al.} respectively. This comparison lends some validity to the fastCAT simulation method. However, the full MC imaging simulations contains assumptions which may need to be modified to return result consistent with experimental data. These assumptions, such as a spatially uniform cone beams and not generating scatter in the primary and secondary collimators, will perhaps degrade the agreement between fastCAT results and experimental measurements. Further work will assess this fidelity and might lead to adjustments in fastCAT so as to attain the highest possible agreement with experimental data.



\subsection{Limitations}



FastCAT suffers from some rigidity in terms of certain parameters. The detector pixel pitch is not a parameter that can be easily modified from the available built-in detectors as this requires rebinning of MC phasespace files in the case of the GOS detector. These files are too large (up to 2.5 GB in size) to include in the software. For a detector like the CWO, modifying the pixel pitch is even more complicated as the detector septa must be moved and the simulation rerun. The implemented scatter correction further restricts the use of fastCAT, as scatter is made for the simulation of a 16-cm diameter water phantom and the agreement between fastCAT and MC simulation will degrade for phantoms that are significantly larger or smaller.

Additionally, in the time comparisons, fastCAT would suffer from memory constraints in completing simulations with high resolution reconstructions, while memory would not be a factor in MC simulations. For example, a simulation of the XCAT head phantom with a phantom size of 1024$\times$1024$\times$10 voxels, 360 projections, and a detector size of 124$\times$512 pixels uses 2.6 GB of RAM. Extending this to a full detector of 1024$\times$1024 pixels and a larger phantom one could easily exceed the RAM capacity of most personal computers. This is due to fastCAT memory requirement scaling linearly with the number of detector pixels. Further work will aim to resolve this rigidity.

Likewise, fastCAT is dependent on the TIGRE python package which in turn has a CUDA dependency. This is a limitation of the application, as many users may not have a CUDA installation, a CUDA capable GPU, or have a GPU at all. Therefore, further work will aim to make a CPU version of fastCAT to increase accessibility.


\subsection{Future applications}

\begin{figure}[h!]
  \begin{center}
  \includegraphics[width=\textwidth,trim={0 0.3cm 0 0},clip]{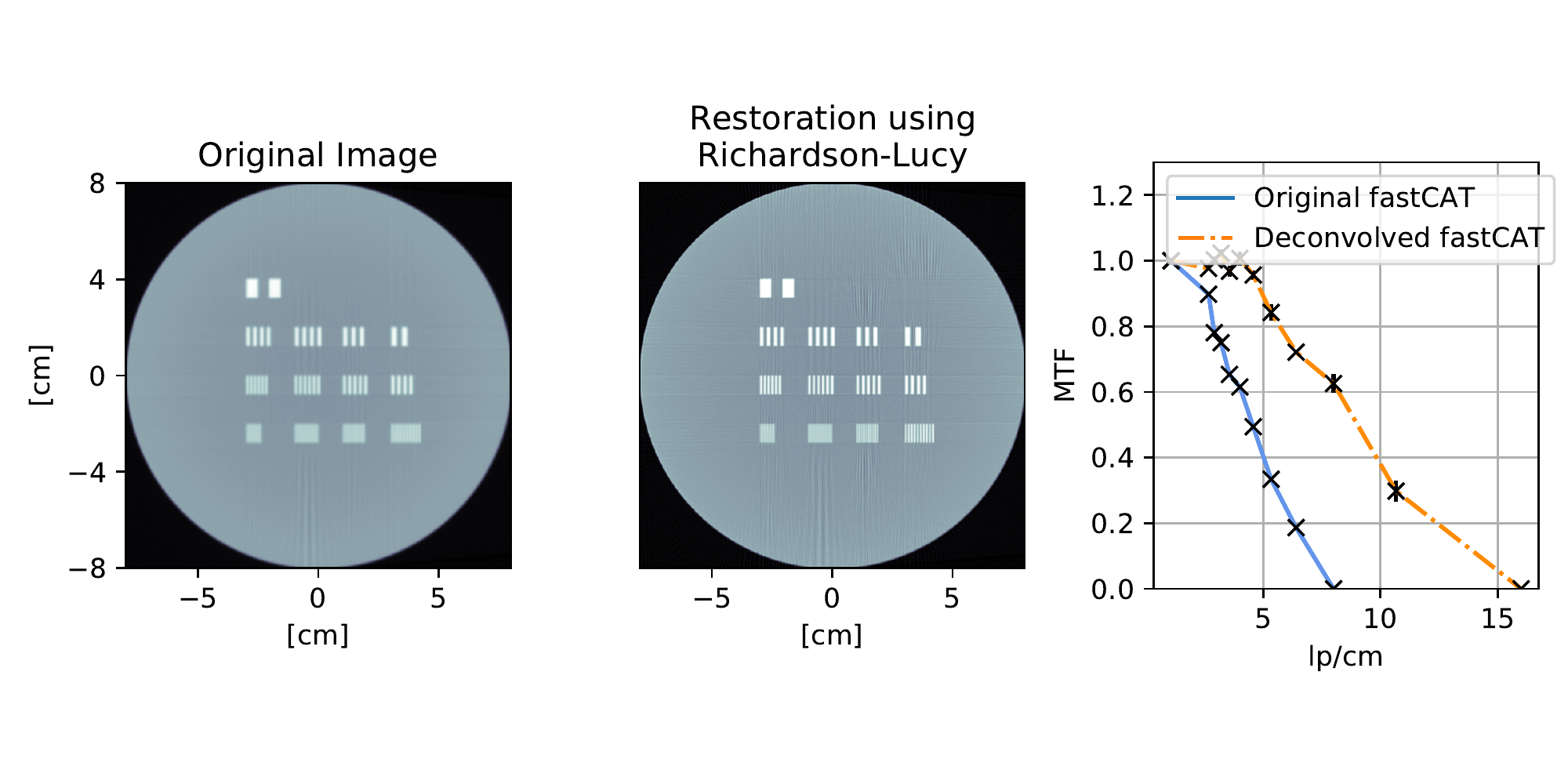}
  \captionv{12}{}
  {The MTF phantom (L) reconstructed conventionally and (Center) using the Richardson-Lucy deconvolution and the PSF calculated by fastCAT. (R) MTF comparison for the two images. The crosses mark the spatial frequencies in the phantom.
  \label{deconvolution} 
    }  
    \end{center}
\end{figure}

Three applications of fastCAT are speculated by the authors: 1) development of new CBCT methodologies (the primary goal of this platform), 2) deconvolution of CT images using an approximation of the PSF, and 3) dataset generation for machine learning (ML) deployment. First, fastCAT can be used to quickly asses the effect of various combinations of beam energies and detector types on CBCT image quality. FastCAT greatly decreases the time it takes to simulate these combinations as it can be done in minutes while giving comparable results to MC in terms of image spatial resolution, contrast and CNR. This allows for a greater flexibility in exploring the parameter space for CBCT imaging equipment and protocols. 
Second, fastCAT generates a PSF for a given beam and detector with an arbitrary focal spot size. If agreement can be achieved between the image MTF of fastCAT and a system experimental data by modification of the focal spot size, one could use the fastCAT PSF to estimate the system PSF. The knowledge of PSF allows one to perform a deconvolution of the image to improve the spatial resolution of CBCT images. In a test case using Richardson-Lucy deconvolution \cite{Richardson1972Bayesian-BasedRestoration,Lucy1974AnDistributions} on a fastCAT MTF phantom (Figure \ref{deconvolution}), CBCT image MTF increased by a factor of two using deconvolution of the projection images. 

Third, fastCAT could be imagined as a tool to link simulation space to real clinical imaging. This could be done by maintaining a fastCAT model of a clinical imaging setup through adjusting user parameters available in fastCAT to agree with monthly quality assurance (QA) images of a Catphan phantom. A clinic could then use fastCAT to generate arbitrary ML training data personalized to a given machine in terms of contrast and MTF. This overcomes some obstacles that arise in terms of ML algorithms which have to be trained on general datasets and can not necessarily respond to changes in medical imaging equipment output over time. This allows for a more personalized approach between dataset generation that links to clinical QA. This method could provide superior performance of ML algorithms and remove risks that ML output would be invalid due to changing machine performance. Automated segmentation of bone CBCT for patient positioning could be suitable for fastCAT, where algorithms could be retrained on adjusted data if drift in QA MTF or contrast indicated a change in image quality.

Additionally, Figure \ref{XCATs} demonstrates the use of fastCAT with an anthropomorphic phantom to simulating more clinically relevant imaging situations. The use of an anthropomorphic phantom with fastCAT allows users to conduct virtual clinical trials\cite{Abadi2020VirtualCOVID-19}. In a virtual clinical trial, different imaging methods could be compared to ascertain the most successful imaging method for a given clinical situation. This could be used, for example, to asses different CBCT acquisition settings for use in radiotherapy treatment planning.

\section{Conclusion}

We presented fastCAT: a fast application for kilovoltage and megavoltage CBCT image generation. FastCAT shows good agreement with measurements with respect to MV beam spatial resolution for a GOS and a CWO detector. The application was also validated with respect to Monte Carlo simulations using contrast modules of a Catphan 515 phantom. A maximum difference of 16 HU between fastCAT and Monte Carlo simulations was observed for the brain, deflated lung, compact and cortical bone, and adipose tissues inserts. The complete fastCAT CBCT dataset generation for a 512$\times$512$\times$10 and a 1024$\times$1024$\times$10 reconstruction size took 40 and 61 seconds, respectively. This is approximately five orders of magnitude faster than the corresponding Monte Carlo simulations.

\section{Acknowledgments}

The authors would like to acknowledge Chelsea Dunning for the contribution to the name fastCAT. We would also like to thank Marios Myronakis for sharing the GOS MTF data with us. This research was enabled in part by support provided by WestGrid (www.westgrid.ca) and Compute Canada Calcul Canada (www.computecanada.ca). The work was partly funded by an NSERC Discovery Grant and the Canada Research Chair program.

\section*{References}
\addcontentsline{toc}{section}{\numberline{}References}
\vspace*{-20mm}





\bibliography{./references}      



\bibliographystyle{./medphy.bst}    


\end{document}